CodaQback: A simplified Python Code facilitating auto-windowing for estimating Seismic Coda attenuation parameter

Nilutpal Bora[1], Rajib Biswas[2*], S. VaasuDevan[3].

[1]*Department of Civil engineering, IIT-Guwahati, Assam, India*

[2]*Geophysical Lab, Department of Physics, Tezpur University, Tezpur-784028, Assam*

[3]*Institute of Remote Sensing, Anna university, Chennai- 600025, TamilNadu.*
*Corresponding Author: rajib@tezu.ernet.in

***Abstract****: Attenuation study of a province is considered as a basic quantity for seismic hazard assessment, ground motion simulation process and source parameter studies. It is already established that the study of two physical processes, first, the seismic sources and second, propagation of the waves, is essential for seismic-hazard mapping, attenuation being one of the properties paying importance to the latter. Here, a computational tool entitled CodaQback is presented. Based on back scattering model, this versatile software is equipped with user friendly graphical user interface. It also allows quick picking of phases for computing coda attenuation parameter. All outputs after each execution step in CodaQback are efficiently exported stepwise into a separate folder in excel and text formats. This CodaQback is checked in real data analysis and there is found to be good matching of computed values with already established ones. It is envisioned that this package will enable user to derive quick and reliable estimation of coda attenuation parameter irrespective of geological and geo-morphological units.*

## 1. Introduction

The energy of seismic wave at several distances from earthquake source is rigorously affected by medium through which wave propagated. Attenuation is one of the vibrant parameters that describe the medium which is mainly a geological medium. It is generally measured by quality factor Q, a dimensionless quantity. By definition, Quality factor gives an estimate of dissipated energy (Aki and Chouet,1975; Knopoff and Hudson, 1964) while spreading through the geological medium. In general Q varies inversely with attenuation. As such, higher attenuation is observed in seismic with lower Q values.

There are several reports concerning estimation of $Q_C$, the quality factor of the coda wave, for different parts of the world. According to these reports, attenuation in crust and lithosphere can be best studied through use of coda wave analysis (Aki and Chouet, 1975; Aki, 1969; Sato,1977; Padhy and Subhadra, 2010; Hazarika et al. ,2009; Biswas et al. a,b, 2013).

The Kopili Fault Zone in Northeast India, is an intraplate earthquake source zone and is characterized as a probable sector for a large forthcoming earthquake. Till now, this fault caused two large earthquake and several intermediate earthquake (Bora et al., 2018). This makes it a possible province for seismic activity as well as socio-economic droughts, and it would be convenient to study attenuation features of this area in order to make it available for other researchers who would be concerned in micro zonation of this active region.

On other hand, Jin and Aki (1988), stated that the regional estimates of $Q_C$ and its spatial nonconformity is directly connected to the seismicity and tectonics, which acts a key role in hazard investigation and engineering seismology. Bora and Biswas, (2017); Bora et al (2017) and Bora et al (2018) investigated the attenuation mechanism for this Kopili region by using their own matlab codes.

During estimation of this vital parameter, most of the researchers resorted to customized code or painstaking analysis and subsequent lengthy computation. It is indeed essential that there should be a robust computational tool which will render this whole analysis effortless but reliable. Keeping this objective in mind, the present work aims at designing software for assessment of the coda-wave attenuation entailing local and regional earthquakes.

In this direction, seismological software are available towards measurement of attenuation of coda waves along with a routine ways. To the best of our knowledge, this is the first endeavor of designing a computational tool for estimating Coda Q based on back scattering model. Consequently, there is no scope of correlating with other existing seismic codes. As for instance, the CodaQ subroutine of SEISAN (Havskov and Ottemoller, 2005) enables the users to attain this parameter. The SGRAPH program (Abdelwahed, 2012) enables the user to estimate the intrinsic and scattering seismic attenuation parameter by using the Multiple Lapse time Window method (Zeng, 1991). Similarly, Predein et al. (2017) recently developed a software package named as CodaNorm, based on coda-normalization method (Knopoff and Hudson, 1964). This package allows the estimation of the seismic quality factor and its frequency dependence ($n$) for direct body waves for different central frequency. In our own developed CodaQback package, the initial data used are:

1. The earthquake waveforms (seismograms) in the PITSA format;

2. The excel file having all the hypocentral factors.

**2. The Single Backscattering model**

We try to measure coda wave attenuation parameter by adopting the single backscattering method proposed by Aki and Chouet (1975). In this model, the amplitude of coda wave ($A_C$) at a central frequency *f* for a preferred frequency band and a precise lapse time *t* measured from the earthquake origin time $t_0$ can be expressed as,

$$A_C(f,t) = S(f)G(f)I(f)\left[t^{-\alpha}e^{\left(-\frac{\pi f t}{Q_C(f)}\right)}\right] \tag{1}$$

where, *S(f)*, *G(f)*, *I(f)* denote the source response, site amplification and instrument response. The geometrical spreading parameter $\alpha$ is measured as unity (constant) in this study as opined by Havskov et al. (1989). According to Havskov et al. (1989), coda waves are considered as backscattered body waves (Aki, 1981 and Aki,1980). The term $Q_C$ denotes the coda wave quality factor which describes the average attenuation of the medium for a predefined area.

The terms *S(f)*, *G(f)*, *I(f)* are time independent, so if we take natural logarithm of the terms *S(f)*, *G(f)*, *I(f)*; then the natural logarithm of the multiplication are time independent, as these terms are time independent. By taking the natural logarithm of equation (1) and rearranging the terms, the eq. can be modified as,

$$\ln[A_C(f,t) \times t] = c - bt \tag{2}$$

This is a simple linear equation where the slope is represented as $b=\frac{\pi f}{Q_C(f)}$ and $c = \ln[S(f)G(f)I(f)]$. Thus $Q_C$ can be estimated from the slope of the aforementioned eq. from the plot of $\ln[A_C(f,t) \times t]$ versus *t*.

### 3. Software Description

The software's primary aim is to provide a user friendly environment rather than the command-line interface. The software also aims at providing the various filter options. With many filters available readily with the software, the user can easily distinguish between the noise and body waves and thereby can select the onset of the waves. Also, the software aims at exporting the output in well-known excel file. Python comes pre-installed with almost all Linux distributions and mac OS.

To install in windows, executables are available in the official website (https://www.python.org). The python version used is 2.7.13. With a slight modification in the code or using the 2to3 converter, it can be made to run with Python3.x. To install additional packages, one must use pip or easy_install or one can also compile from the source. The wheels for the packages can be found at http://www.lfd.uci.edu/~gohlke/pythonlibs/ and to install from the command line using pip is shown below.

**pip install matplotlib numpy scipy xlwt**

**4. Data Processing**

The input files for the software are PITSA (PITSA: Programmable Interactive Toolbox for Seismological Analysis) files. Internally, PITSA uses a data format designed towards the needs of earthquake seismology. The PITSA files can be obtained by using the SEIPITSA converter which comes with the SEISAN software (EarthQuake Analysis Software).

For example :

Let us consider a seisan file 2003_091_14_49_58_00928_1.SEI.

We can convert the .SEI to PITSA files using the converter. When converted successfully we will have three PITSA files corresponding to three components likely Z, NS and EW.

Pitsa001.001 corresponds to Z component (particle), .002 to NS and .003 to EW component. Only the file with extension .001 is required for the coda software to process the data for that particular earthquake. For simplicity, only the vertical components are used to study and estimate the Coda Q value. All these PITSA files for many sites should be put together in the folder named PITSA. All the files will be read and after that we proceed to the next step. If no PITSA files are found, the software will exit at this stage.

The algorithm of the program is shown in figure. 1.

**4.1. EXECUTION**

**4.1.1 PITSA TO TXT**

Coda Normalisation software looks for the PITSA folder. Then the software looks for the PITSA files in the directory (files with extensions .001). If read successfully, the software creates a directory called "Data" and converts all these files into file which is understandable by the software and saves it in that Data directory. Software names the files as 1.txt for pitsa001.001, 2.txt for pitsa002.001 and so on. Those files should be left undisturbed when the software is running.

**4.1.2 GUI CREATION AND PLOTTING THE VALUES**

CodaQback software looks for the Data folder and read the first file (1.txt). It then initiates the GUI (Graphical User Interface) window and plots the values. The GUI window is shown in the (Figure 2). Each seismogram was bandpass-filtered at five frequency bands (1-2; 2-5; 4-8; 6-12; 9-15 Hz) with central frequencies at 1.5, 3.5, 6, 9 and 12 Hz. All these five butterworth filter options are available in the top – right corner of the software which can be used to visualize the graphs in that particular frequency. There are three entry fields available. First one for the P-Wave velocity and S-Wave velocity (in km/s) and third one for the Coda-wave moving window time (in seconds). The user must insert the time in seconds (Eg. 1.26s, 5s, 10s) and select the onset of P wave and S wave manually.

After the P and S waves are selected, the user should press the Execute button present in the toolbar. Once execute button is pressed, the core part of the software starts to execute. It will take the onset P-waves and onset S-waves and computes the origin time ($t_0$) by using the following equation

$$t_0 = t_P - \frac{t_S - t_P}{\frac{v_P}{v_S} - 1}$$

Once the Origin time is computed, Onset of coda wave is computed using

$$t_0 + 2(t_S - t_0)$$

Then, the coda wave will be drawn from 30 seconds to 90 seconds with increment of 10 seconds. The lapse time windows of the coda wave used in this software are started from twice of the S-wave travel time so that there was no influence of direct S-waves. While there is usually no bound for the maximum spans of the coda window length, but in our case, we have found that the SNR (Signal to Noise ratio) values are more than 2 for most of data sets for 100 seconds window. Thus the upper limit has been set to 90 seconds. Those values are shaded in the graph once the execute button is pressed (Figure 3a and 3b).

After execution, the user can see the $t_0$ values using the "Show Values" button present in the toolbar. Execution will create a directory called "Info" and then creates another directory with the file name (Eg: 10.txt). Here 10 represent the serial number of the analyzing event. Inside that directory, the software saves a .png copy for all the coda windows with the file name (10.txt_allWindows.png).

Next, the user has to press the Compute button. Once the button is pressed, the coda amplitude will be computed for every seconds given in the entry field (Coda moving Window time). So if, 2.56 seconds was given, the software computes RMS amplitudes of each 2.56s window length. Then, the moving window slides along the coda window with steps of half of the moving window length (i.e., 1.28s in this case) and again computes RMS amplitudes of each 2.56s window length. Thus for 30s window, we will have 22 Coda wave RMS amplitude values.

Once the Ac values is computed for 30-sec, a dialog box will appear stating "Success-Ac" with the corresponding names as shown (Figure 4a and 4b). Once OK button is pressed, it will compute for the next window (i.e., for 40 sec window length).

Also, on the top-right corner we can see the successful states of the *software's execution*. A screenshot of the software is shown in figure 4(c). The software saves all the files with their corresponding names in the folder that was created with the file name. For butterworth filter, *butter function* from the *Scipy.signal* package is used and for RMS values, the *sqrt function* from *numpy* is used. Once the Ac values are computed, the software then computes the $Q_C$ values

The following table lists the maximum amplitude values ($A_C$) saved in 40-secWindow.txt which is saved in the 10.txt folder. Similar to this, each file will have their corresponding values.

Once the software computes the maximum amplitudes for all the seismograms (after 90-sec window), it starts to compute the $Q_C$ values. For the computation, it takes the Ac values from the file created and saved in the previous step. First the software opens the 30-sec window file, reads all the smoothed values for the first frequency, computes the slope and intercept value and then plots the graph. Thus $Q_C$ can be obtained from the slope of the straight line of equation 2 at each central frequency. Towards estimation of $Q_C$, we have found few undesirable values which are excluded (Woodgold, 1994). Once the $Q_C$ values are computed for 30-sec, a dialog box will appear stating "*Success-Qc*" with the corresponding names as shown in Figure 4(d) and 4(e). Once OK button is pressed, it will compute for the next window.

Also, on the top-right corner we can see the successful states of the software's execution. A screenshot of the software is shown in Figure 4(f). Once the software computes all the $Q_C$ values and also its corresponding graph, it goes to the next window and starts repeating the process again for the next window. For computing the slope and intercept values, *scipy.stats.linregress* method is used. It is a linear regression technique used to compute those values. The computed values for 30-sec and 90-sec window is tabulated below (Table 2).

(Figure 5a) shows the slope graph for 30-sec Window and (Figure 5b) shows the slope graph for 90-sec window (for the values shown above).

After all the process for the first PITSA file (earthquake event) is over, the GUI quits and goes to the Execution part. Another GUI window will pop up for the second file and the user should select the onset of P and S wave and should press the Execute and Compute button.

For convenience of the output of large data volume, a file in the format of Excel at the output of the program is created (Final.xls), in which a separate worksheet is generated for each central frequency, where all normalized amplitudes for all windows with the corresponding names are also entered. The calculation results are saved in database, the final calculation is made by means of Microsoft Excel.

**5. Impact and test with real data**

The software package CodaQback was tested by us to evaluate the coda wave attenuation in the Kopili fault zone. In order to do so, we have analyzed almost 300 digital seismograms of earthquakes within the range 2.1-3.9($M_L$), recorded by six stations (Table 1) surrounded by an area with epicentral distance ~180 km. The value of $Q_C$ for one earthquake event recorded at station RUP as obtained by using this python package is shown in Figure 5 (a) and (b); and table 2. The analysis of attained coda wave attenuation parameters revealed that:

1. The estimated values of $Q_C$ revealed that (table 2, figure 5(a) and (b)), for window length of *30sec*, the value of $Q_C$ increases from *149* at frequency *1.5Hz* to *1685* at frequency *12Hz*. Similarly, an increasing fashion in $Q_C$ is observed for all the lapse time window up to 90sec.

2. Our estimated frequency dependent relations for Kopili area comes out to be as *(65±3)f$^{(1.29±0.09)}$* for 30 sec window and *(217±7)f$^{(0.93±0.02)}$* for *90 sec* window. Which are almost similar to the result obtained by Bora et al. (2018). They reported that frequency dependent relations for Kopili fault were *(63±6)f$^{(1.33±0.11)}$* for *30 sec* window and *(213±5)f$^{(0.91±0.03)}$* for *90 sec* window length. Although, there is marginal difference between the estimates, this can be attributed to the manual and automated picking of the coda wave onsets.

3. The higher values of *n* suggest that the entire area is highly heterogeneous in nature, which was also supported by our previous studies (Bora et al. (2018)

## 6. Conclusions:

Conclusively, we have devised CodaQback- a versatile computational tool which we utilized in Kopili for deciphering coda Q. Based on back scattering S-coda envelops, the computational package offers effortless estimation of this vital parameter. The attained estimates through this package are excellently tallying with our published results which were derived through painstaking analysis. As it is written in Python, there is ample scope of incorporation of other routine codes which makes it a very promising computational tool. As the authors have utilized and verified for real data, it is envisioned that this developed tool will prove beneficial to seismologists, geoscientists working in these relevant fields providing a tunable platform.

**Declaration of Conflict of Interest:**

On behalf of all authors, the corresponding author states that there is no conflict of interest.

**REFERENCE**

Abdelwahed M F. 2012. "SGRAPH (SeismoGRAPHer): Seismic waveform analysis and integrated tools in seismology." *Computers & Geosciences,* 40:153–65. http://dx.doi.org/10.1016/j.cageo.2011.06.019

Aki, K. 1969. "Analysis of the Seismic Coda of Local Earthquakes as Scattered Waves." 74(2):615–31.

Aki, K. 1980. "Attenuation of Shear-Waves in the Lithosphere for Frequencies from 0.05 to 25 Hz." *Physics of the Earth and Planetary Interiors* 21(1):50–60.

Aki, K. 1981. "Source And Scattering Effects On The Spectra Of Small Local Earthquakes." 71(6):1687–1700.

Aki, K., & Chouet, B. 1975. "Origin of Coda Waves: Source, Attenuation, and Scattering Effects." *Journal of Geophysical Research* 80(23):3322.

Biswas, R., Baruah, S., Bora, D.K., Kalita, A., Baruah, S., 2013a. The effects of attenuation and site on the spectra of microearthquakes in the Shillong region of Northeast India. Pure Appl. Geophys. 170 (11), 1833–1848. http://dx.doi.org/10.1007/s00024-012-0631-0

Biswas, R., Baruah, S., Bora, D., 2013b. Influnce of attenuation and site on microearthquakes' spectra in Shillong Region of Northeast India: a case study. Acta Geophysica 61, 886–904.

Bora, N., & Biswas, R. 2017. "Quantifying Regional Body Wave Attenuation in a Seismic Prone Zone of Northeast India." *Pure and Applied Geophysics.,* 174(5): 1953-63.

Bora, N., Biswas, R., & Bora, D. 2017, "Assessing attenuation characteristics prevailing in a seismic prone area of NER India". *J. Geophys. Eng.* **14**:1368-1381, DOI: https://doi.org/10.1088/1742-2140/aa7d11

Bora, N., Biswas, R., & Dobrynina, A. A. 2018, "Regional variation of coda Q in Kopili fault zone of northeast India and its implications". *Tectonophysics*, 722:235-248, https://doi.org/10.1016/j.tecto.2017.11.008.

Havskov, J., Malone, S., Mcclurg, D. & Crosson, R. 1989. "Coda-Q for the State of Washington." *Bulletin Of The Seismological Society Of America* 79(4):1024–38.

Havskov, J., Ottemoller, L., 2005. SEISAN (Version 8.1): The Earthquake Analysis Software for Windows, Solaris, Linux, and Mac OSX Version 8.0, pp 254.

Hazarika, D., Baruah, S. & Gogoi, N. K. 2009. "Attenuation of Coda Waves in the Northeastern Region of India." *Journal of Seismology* 13(1):141–60.

Jin, A., & Aki, K. 1988. "Spatial and Temporal Correlation between Coda Q and Seismicity in China." *Bulletin of the Seismological Society of America* 78(2):741–69.

Padhy, S., & Subhadra, N. 2010. "Attenuation of High-Frequency Seismic Waves in Northeast India." *Geophysical Journal International* 181(1):453–67.

Predein, P. A., Dobrynina, A. A., Tubanov , T. A. and German, E. I. 2017. "CodaNorm: A software package for the body-wave attenuation calculation by the coda-normalization method", *SoftwareX*, Volume 6, 30-35

SATO, H. 1977. "Energy Propagation Including Scattering Effects Single Isotropic Scattering Approximation." *Journal of Physics of the Earth* 25(1):27–41.

Woodgold, C. R. D. 1994. "Coda Q in the Charlevoix , Quebec , Region : Lapse-Time Dependence and Spatial and Temporal Comparisons." *Bulletin of the Seismological Society of America* 84(4):1123–31.

Zeng Y. 1991. "Compact solutions for multiple scattered wave energy in time domain." *Bulletin of the Seismological Society of America.,* 81:1022–9.

**List of Figures:**

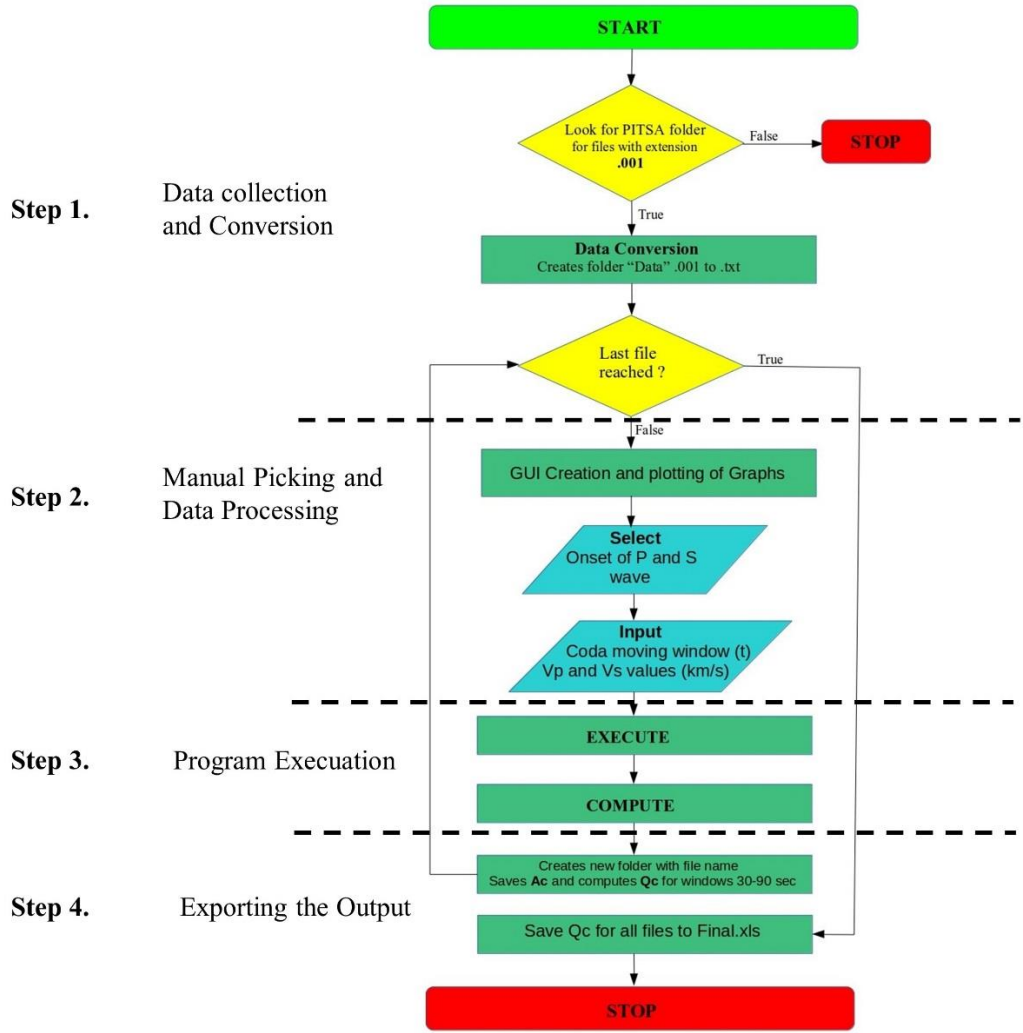

**Figure 1**: Flowchart of the software package CodaQback.

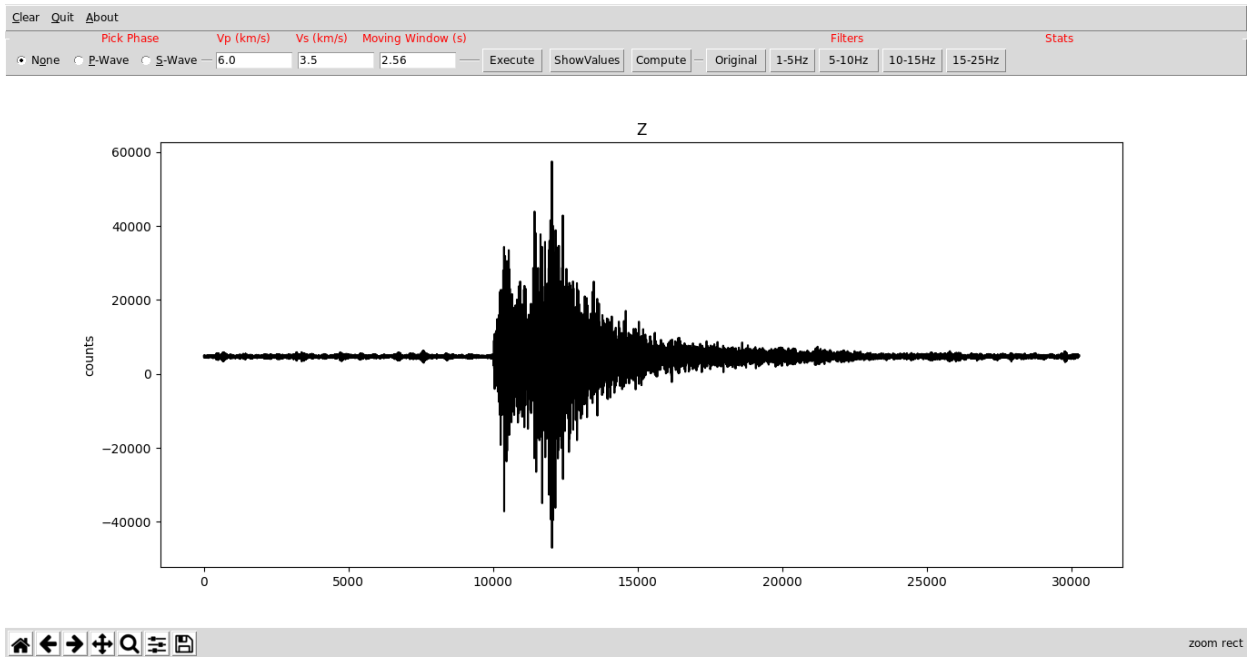

**Figure 2**. The main GUI window of Codaqback showing original vertical component of non-filtered seismogram.

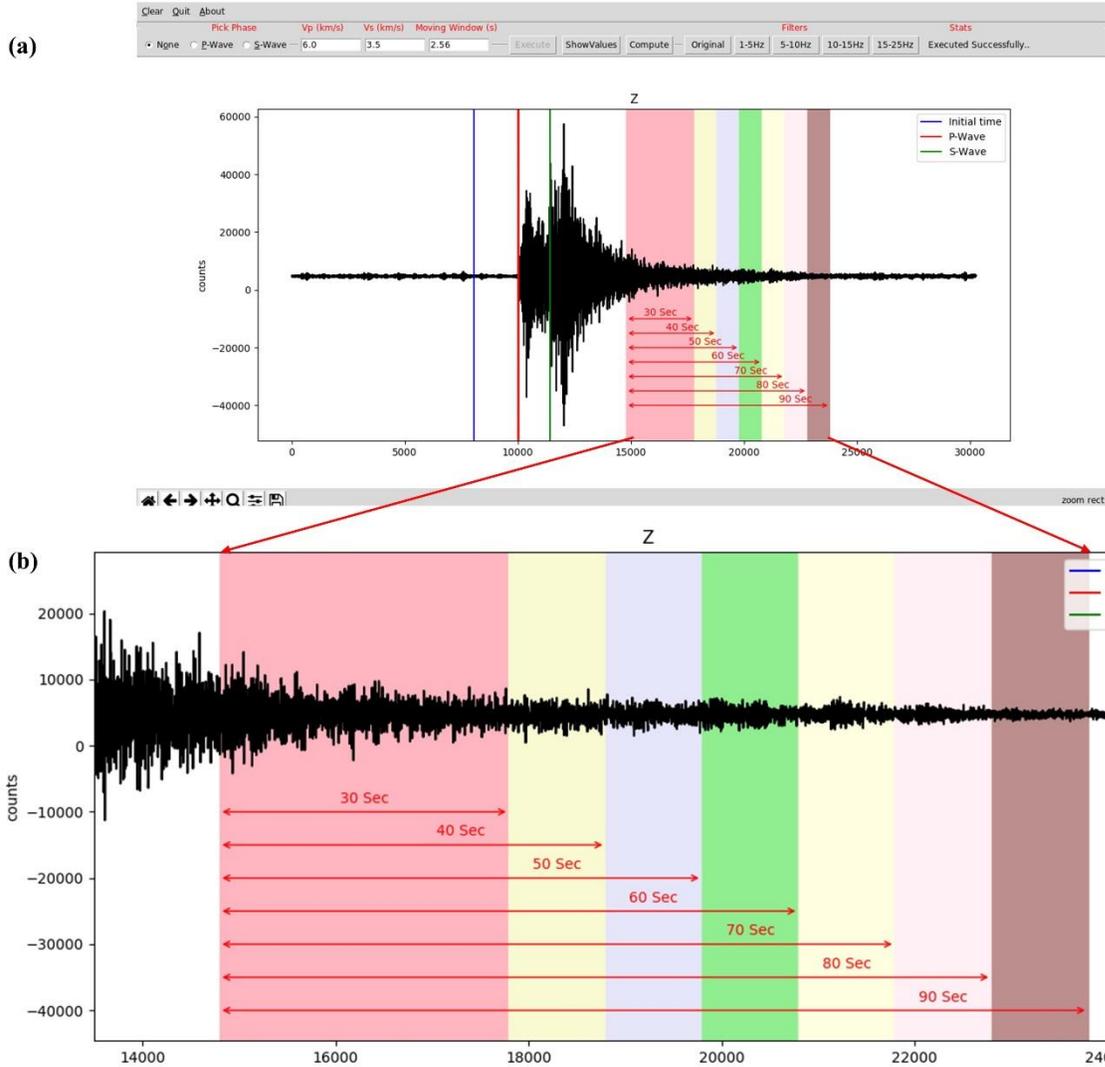

**Figure 3.** The example of earthquake waveform processing. (a) Red line and green line represent the onset of P and S wave (manually selected). Blue line show the origin time of the earthquake. (b) The colored portion in the tail portion of the seismogram shows the lapse time window for coda wave, for which calculations of seismic quality factor were carried out, starting from 30second to 90second.

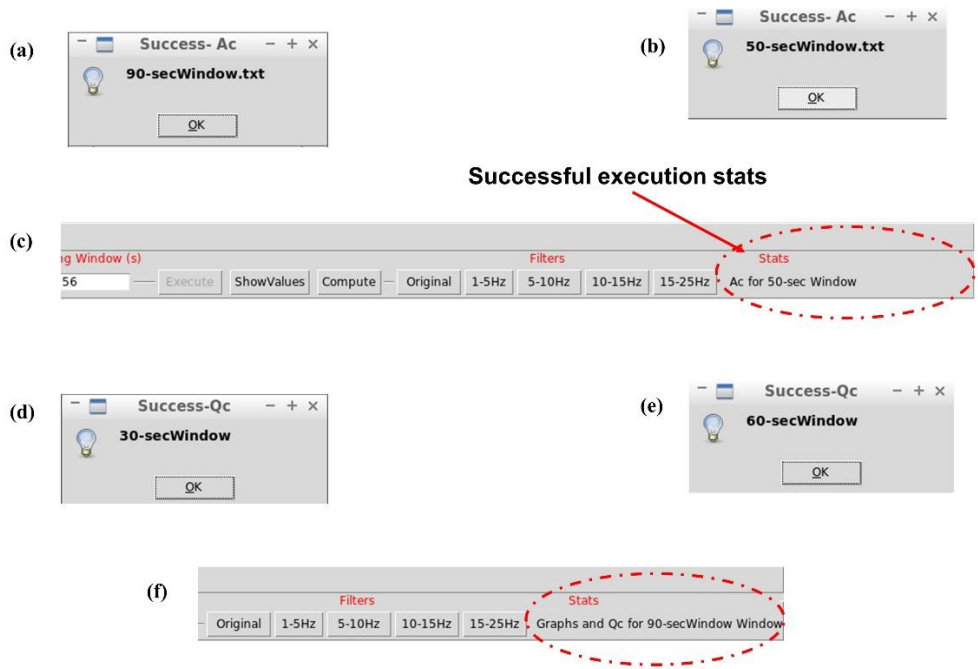

**Figure 4.** Screen shots of the software execution steps. (a) And (b) shows the successful calculation of coda wave Amplitude ($A_C$) corresponding to their window length. (c) Shows the successful execution states (inside the red circle. (d) And (e) show the pop-up window showing the successful execution of $Q_C$ corresponding to their window length. (f) Shows the successful execution states of $Q_C$ and indicating the pop-up windows represent the estimated $Q_C$ values for corresponding window lengths.

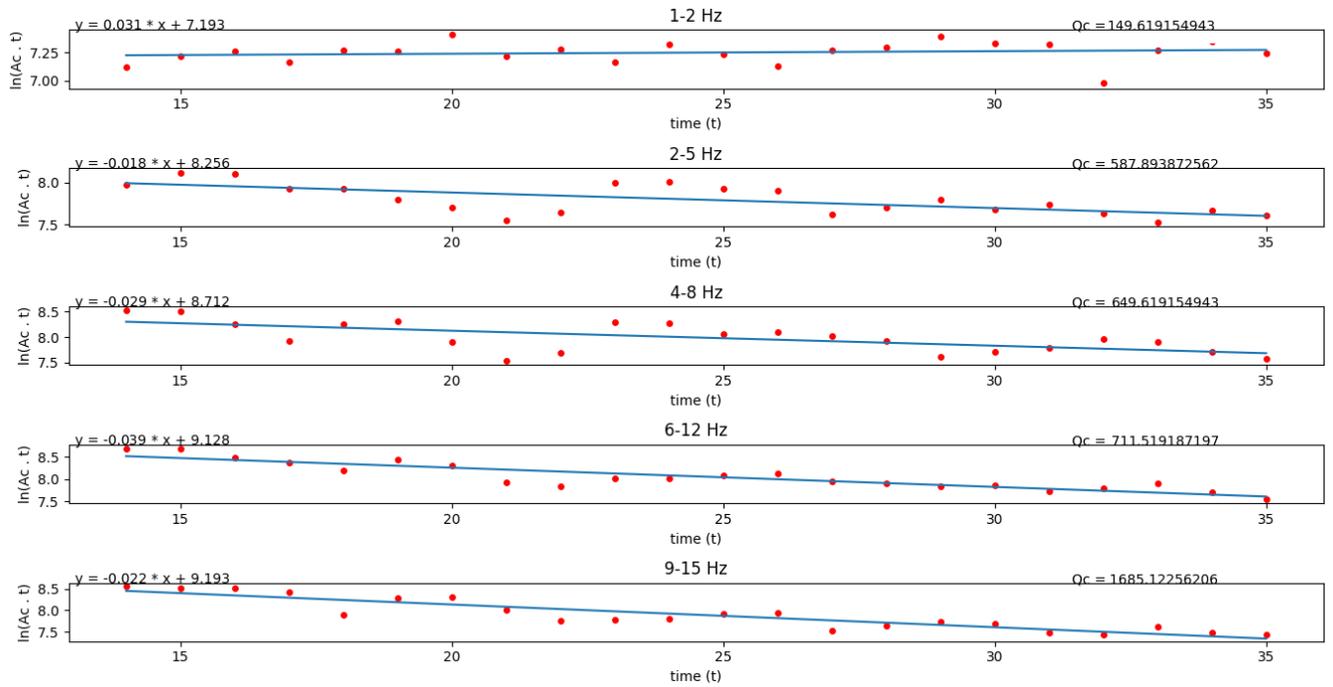

**Figure 5(a).** Screen shots of the software execution steps. The value of $Q_C$ for one earthquake event recorded at station RUP as obtained by using this python package corresponding to window length of 30 second.

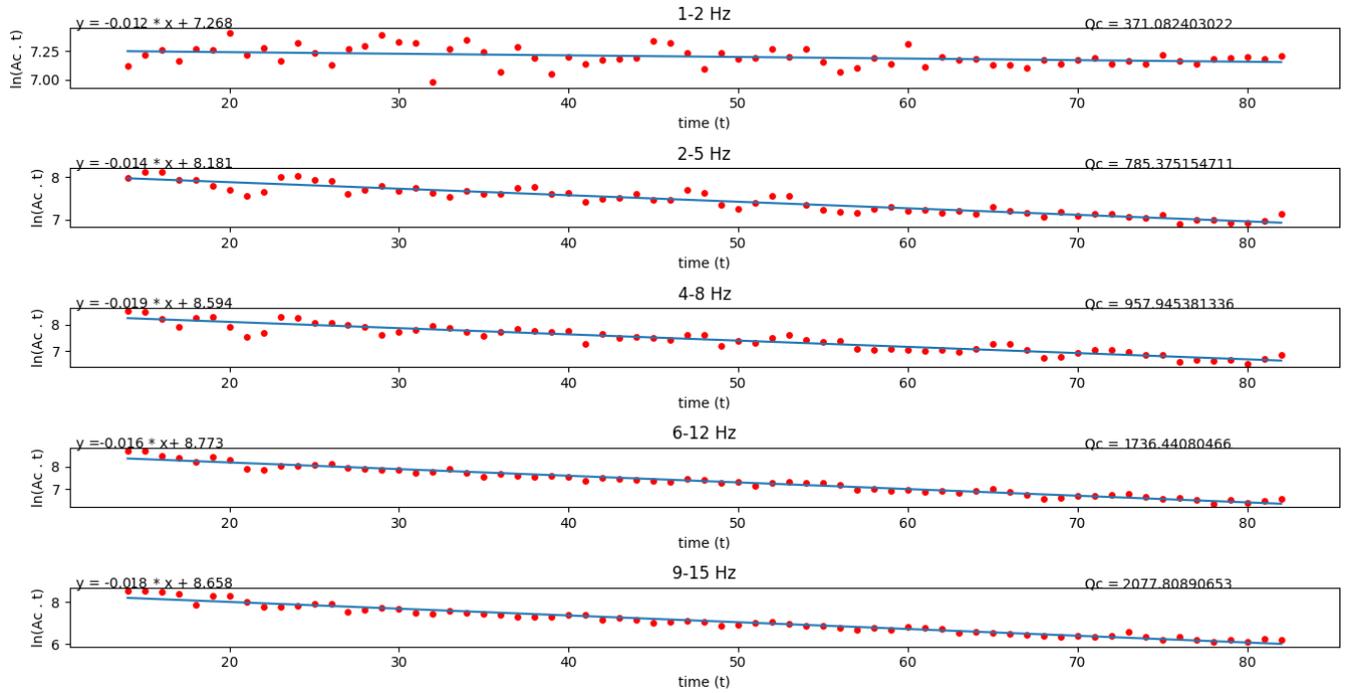

**Figure 5(b).** Screen shots of the software execution steps. The value of $Q_C$ for one earthquake event recorded at station RUP as obtained by using this python package corresponding to window length of 90 second.

**Lists of Tables**

Table 1: Location of six stations

Table 2: The value of $Q_C$ for one earthquake event recorded at station RUP as obtained by using this python package

Table 1.

| Station Name | Latitude (Degrees) | Longitude (Degrees) | Elevation (Meter) | Ground Type |
|---|---|---|---|---|
| BKD | 26.89 | 92.11 | 210 | Hard Rock |
| BPG | 26.99 | 92.67 | 130 | Hard Rock |
| DMK | 26.21 | 93.06 | 200 | Alluvium |
| RUPA | 27.20 | 92.40 | 1470 | Hard Rock |
| SJA | 26.93 | 92.99 | 150 | Hard Rock |
| TZR | 26.61 | 92.78 | 140 | Hard Rock |

Table 2:

| For 30Sec window length | | For 90Sec window length | |
|---|---|---|---|
| Frequency (Hz) | $Q_C$-values | Frequency (Hz) | $Q_C$-values |
| 1.5 | 149 | 1.5 | 371 |
| 3.5 | 587 | 3.5 | 785 |
| 6 | 649 | 6 | 957 |
| 9 | 711 | 9 | 1736 |
| 12 | 1685 | 12 | 2077 |